\documentclass[12pt]{article}
\usepackage{}
\usepackage{geometry}             % See geometry.pdf to learn the layout options. There are lots.
\usepackage{graphicx}
\usepackage{amssymb}
\usepackage{amsmath}
\usepackage{epstopdf}
\usepackage{comment}
\usepackage{cite}
\usepackage{caption}
\usepackage{arcs}

\usepackage[hyperindex=true,
          pdfstartview=FitH,
          bookmarksnumbered=true,
          bookmarksopen=true,
          citecolor=blue,
          linkcolor=blue,
          colorlinks=true,
          pdfborder=001,
          unicode]{hyperref}

\allowdisplaybreaks

\begin{document}
\title{Critical phenomena of static charged
AdS \\black holes in conformal gravity}
\author{Wei Xu$^{1,2}$ and
Liu Zhao$^{2}$\\
$^{1}$School of Physics, Huazhong University of Science and Technology, \\ Wuhan 430074, China\\
$^{2}$School of Physics, Nankai University, Tianjin 300071, China\\
{\em email}:
\href{mailto:xuweifuture@gmail.com}{xuweifuture@gmail.com}
and
\href{mailto:lzhao@nankai.edu.cn}{lzhao@nankai.edu.cn}}
\date{}
\maketitle
\begin{abstract}
The extended thermodynamics of static charged AdS black holes in conformal
gravity is analyzed. The $P-V$ criticality of these black holes has some unusual
features. There exists a single critical point with critical temperature $T_c$
and critical pressure $P_c$. At fixed $T>T_c$ (or at fixed $P>P_c$), there are
two zeroth order phase transition points but no first order phase transition
points. The systems favors large pressure states at constant $T$, or
high temperature states at constant $P$.
\end{abstract}

\section{Introduction}
Recently the study of thermodynamics of black holes in AdS spacetime has been
generalized to the extended phase space, in which the cosmological constant is
identified as a thermodynamic pressure and its variations are included in the
first law of black hole thermodynamics \cite{Kastor:2009wy,Dolan:2010ha,
Dolan:2011xt,Dolan:2011jm,Cvetic:2010jb}. Such studies are mainly motivated
by the geometric derivation of the Smarr relation \cite{Kastor:2009wy,
Dolan:2011xt,Cvetic:2010jb,Dolan:2013ft}, in which a term consisting of the
pressure and its conjugate ``thermodynamic volume'' appears. Once one takes the
cosmological constant as thermodynamic pressure in the first law, the black hole
mass should be explained as enthalpy rather than internal energy of the system.
Besides, the criticality associated with the pressure was discussed in the
extended phase space \cite{Kubiznak:2012wp,Gunasekaran:2012dq,
Dolan:2012jh,Belhaj:2012bg,Hendi:2012um,Chen:2013ce,Zhao:2013oza,
Belhaj:2013ioa,Poshteh:2013pba,Altamirano:2013ane,Cai:2013qga,
Belhaj:2013cva,Altamirano:2013uqa,Zou:2013owa,Altamirano:2014tva,
Wei:2014hba,Kubiznak:2014zwa,Zou:2014mha,Xu:2013zea}. With the extended
structure of the thermodynamic phase space, one can often find
phase transitions and critical points of various kinds.
For instance, an interesting phenomena of black hole reentrant phase
transition is found recently in four-dimensional Born-Infeld-AdS black hole
\cite{Gunasekaran:2012dq}, higher dimensional rotating AdS black holes
\cite{Altamirano:2013ane}, Kerr AdS black holes \cite{Altamirano:2013uqa},
GB-AdS black hole \cite{Wei:2014hba} and the third order Lovelock AdS
black holes \cite{Xu:2014tja}, which is observed previously in
multi-component fluids \cite{Narayanan}. It is also possible to
take other parameters as novel dimensions of the thermodynamic phase space,
e.g. the Born-Infeld parameter in the Born-Infeld-AdS black holes
\cite{Gunasekaran:2012dq} and the GB parameter in GB-AdS black holes
\cite{Xu:2013zea}.

The reentrant phase transition includes both first and zeroth order phase
transitions. However, the origin of zeroth order phase transition is still
unclear. This makes it interesting to consider black hole systems
containing only the zeroth order phase transition. The thermodynamics of the AdS
black holes in conformal gravity \cite{Li:2012gh} to be studied in this
work provides precisely such an example.

On the other hand, although taking the cosmological constant
as a thermodynamic pressure is nowadays a common practice, such operations
implicitly assume that gravitational theories which differ only in the values
of the cosmological constants are considered to fall in the ``same class'',
with unified thermodynamic relations. The common excuse for doing this is
that the classical theory of gravity may be an effective theory which follows
from a yet unknown fundamental theory, in which all the presently ``physical
constants" are actually moduli parameters that can run from place to place in
the moduli space of the fundamental theory. Given that the fundamental theory is
yet unknown, it is preferable
to consider the extended thermodynamics of gravitational theories involving
only a single action, which requires that all variables included in the
thermodynamical relations must be integration constants. It so happens that,
in conformal gravity, the cosmological constant comes as an integration constant
rather than as a bare parameter which appear explicitly in the classical action.

In this paper, we consider the $P-V$ criticality of the static charged AdS Black
hole in conformal gravity. There exists a single critical temperature,
above which there are two zeroth order phase transition points.
In the next section, we first revisit the thermodynamics of static charged AdS
black hole in conformal gravity in the extended phase space. Section 3 is
devoted to the $P-V$ criticality, particular attention is paid toward the
appearance of critical points and the zeroth order phase transitions. Finally,
some concluding remarks are given in the last section.

\section{Thermodynamics of static charged AdS Black hole in conformal gravity
revisited}

We consider the static charged AdS Black hole in conformal gravity.
The action of conformal gravity reads
\begin{align}
S=\alpha\int d^4x \sqrt{-g}\left(\frac{1}{2}C^{\mu\nu\rho\sigma}
C_{\mu\nu\rho\sigma} +\frac{1}{3} F^{\mu\nu}F_{\mu\nu}\right),
\label{action}
\end{align}
where the unusual sign in front of the Maxwell term is inspired by critical
gravity \cite{Lu:2011zk} and is required by requiring that the Einstein
gravity emerges from conformal gravity in the infrared limit
\cite{Maldacena:2011mk}.

The static black hole solution with AdS asymptotics for conformal gravity
is found in \cite{Li:2012gh}, which takes the form
\begin{align}
  &\mathrm{d} s^2=-f(r)\mathrm{d}t^2+\frac{\mathrm{d}r^2}{f(r)}+r^2\mathrm{d}
  \Omega_{2,\epsilon}^2, \label{metric}
\end{align}
where
\begin{align}
  f(r)=-\frac{1}{3}\Lambda r^2+c_1 r+c_0+\frac{d}{r},
\end{align}
and
\begin{align}
  A=-\frac{Q}{r}\mathrm{d}t
\end{align}
is the corresponding Maxwell field. $\mathrm{d}\Omega_{2,\epsilon}^2$
represents the line element of a $2d$ maximally symmetric Einstein space
with constant curvature $2\epsilon$, with $\epsilon = 1, 0$ and $-1$,
respectively. There are six parameters $Q, c_0, c_1, d, \Lambda, \epsilon$ in
the solution, five of which are integration constants, and the last one
determines the spacial sectional geometry of the horizon. These parameters
obey a constraint
\begin{align}
  3c_1d+\epsilon^2+Q^2=c_0^2,
\label{relation}
\end{align}
so there are actually only 5 independent parameters ($c_0$ is to be
considered to be determined by other parameters as in \eqref{relation}).
Except the discrete parameter $\epsilon$, the rest 4 parameters
$Q, c_1, d,\Lambda$ are related to conserved charges: electric charge, charge of
massive spin-2 hair, enthalpy and pressure, respectively. However, at
fixed charges $Q, c_1, d$, $\Lambda$ and $\epsilon$, there still exist a
discrete freedom in choosing the integration constant $c_0$:
\begin{align*}
c_0=\pm\sqrt{3c_1d+\epsilon^2+Q^2}.
\end{align*}

Under the point of view of taking the cosmological constant $\Lambda$ as a
thermodynamic pressure
\begin{align}
  P=-\frac{\Lambda}{8\pi},
\end{align}
the energy calculated by employing the Noether charge associated with the
time-like Killing vector \cite{Lu:2012xu} should be identified with the
enthalpy $H$ of the gravitational system. It reads
\begin{align}
  H&=\frac{\alpha(c_0-\epsilon)(\Lambda r_0^2-3c_0)}{72\pi r_0}
  +\frac{\alpha(2\Lambda r_0^2-c_0+\epsilon)d}{24\pi r_0^2},\nonumber\\
  &=\,{\frac { \alpha\left( c_{{1}}c_{{0}}-c_{{1}}\epsilon-16\pi P\,d
 \right) }{24\pi }},
\end{align}
where $r_0 > 0$ denotes the largest real root of $f(r)$ which corresponds to
the event horizon black hole\footnote{We take $\Lambda<0$, and so there is no
cosmological horizon in the solution.}, $\alpha$ is overall coupling which is
present in the action \eqref{action}.
Because of the double-valuedness of the integration constant $c_0$, one
immediately sees a double-valued behavior of the enthalpy. Such behaviors are
also present in the four dimensional charged rotating black hole
\cite{Liu:2012xn} and six dimensional static black holes \cite{Lu:2013hx}
of conformal gravity. One can expect that the temperature and Gibbs free energy
may also be double-valued. These double-valued variables must be all considered
in order to have a holistic look at the thermodynamics in the extended phase
space of the black hole.

The thermodynamical conjugate of the pressure, i.e. the ``thermodynamic
volume'', is given by
\begin{align}
  V=\left(\frac{\partial H}{\partial P}\right)_{S,Q_{{e}},\Xi}
  =-\,{\frac {\alpha\,d}{3 }}.
\end{align}
It can be seen that the sign of $V$ is determined by the sign of the parameter
$d$. Besides $P$ and $V$, all the other thermodynamic quantities are given in
\cite{Li:2012gh}. The temperature is
\begin{align}
  T=\,{\frac {8\pi P\,r_0^{3}-3\,c_{{0}}r_{{0}}-6\,d}{12\pi \,{r
_{{0}}}^{2}}},
\label{temperature}
\end{align}
and its conjugate, i.e. the entropy is
\begin{align}
  S=\,{\frac {\alpha\,\left(\epsilon r_0-c_{{0}}r_{{0}}-\,  3\,
d \right) }{6r_{{0}}}}.
\end{align}
The electric charge and the conjugate potential are respectively
\begin{align}
  &Q_{{e}}=\,{\frac {\alpha\,Q}{12\pi }},\\
  &\Phi=-{\frac {Q}{r_{{0}}}}.
\end{align}
The parameter $c_1$ is a massive spin-2 hair which is now taken as a novel
dimension in the thermodynamic phase space. We label this novel dimension and its conjugate as $\Xi, \Psi$:
\begin{align}
  &\Xi=c_{{1}},\\
  &\Psi=\,{\frac {\alpha\, \left( c_{{0}}-\epsilon \right) }{24\pi }}.
\end{align}
Throughout this work, we will take the normalization $\alpha=2$.
It can be checked that the first law of thermodynamics
\begin{align}
  {\it \mathrm{d}H}=T{\it \mathrm{d}S}+\Phi\,{\it \mathrm{d}Q}_{{e}}+ \Psi\,
  \mathrm{d}\Xi + V\,\mathrm{d}P
\end{align}
and the Smarr relation
\begin{align}
  H=2PV+\Psi\,\Xi \label{Smarr}
\end{align}
hold in the extended thermodynamic phase space \cite{Li:2012gh}. The absence of $TS$ and
$\Phi Q$ terms in the Smarr relation can be explained by scaling arguments.
The enthalpy can be viewed as a homogeneous function of the extensive variables
$S, Q_e, \Xi, P$, i.e.
\begin{align*}
H=H(S, Q_e, \Xi, P).
\end{align*}
Assuming each extensive variable has a scaling dimension which is denoted
$d_S, d_Q, d_\Xi, d_P$ respectively. If the enthalpy
$H$ itself has scaling dimension $d_H$, then after a rescaling of the extensive
variables we get
\[
\lambda^{d_H} H =H(\lambda^{d_S} S, \lambda^{d_Q} Q_e, \lambda^{d_\Xi} \Xi,
\lambda^{d_P} P).
\]
Taking the first derivative with respect to $\lambda$ and then setting
$\lambda=1$, we get
\[
{d_H} H ={d_S} T S+ {d_Q} \Phi Q_e + {d_\Xi} \Psi \Xi
+ {d_P} V P.
\]
After a little calculation, one can find that $c_1$ scales as [length]$^{-1}$;
$c_0$ scales as [length]$^0$; $\Lambda$ scales as [length]$^{-2}$ which is the
same with $P$; $d$ scales as [length]$^1$; $Q$ scales as [length]$^0$, which is
the same with $Q_{e}$; $S$ scales as [length]$^0$; $H$ scales as
[length]$^{-1}$; $d$ scales as [length]$^1$, which results in the Smarr relation
\eqref{Smarr}, as expected.

While considering critical behaviors, the Gibbs free energy will play an
important role. It is given as follows:
\begin{align}
G&=H-TS\nonumber\\
   &=-\frac{\alpha Pd}{3}
   -\frac{\alpha}{24\pi \,r_0^{3}} \bigg(\,{ \left( c_{{0}}r_{{0}}
   -\epsilon r_{{0}}+3\,d \right)
 \left( c_{{0}}r_{{0}}+2\,d \right) }+\,{ \left( c_{{0}}-\epsilon \right)
 \left( c_{{0}}r_{{0}}+
d \right)r_0} \bigg) . \label{Gibbs1}
\end{align}
On the other hand, the Helmholtz free-energy can be obtained from the
Euclideanized action:
\begin{align}
  F=\,{\frac {\alpha\, \bigg( 2\, \left( c_{{0}}-\epsilon \right)
\epsilon\,r_{{0}}+ \left( 3\,\epsilon+8\pi P\,r_0^{2} \right) d
 \bigg) }{24\pi \,r_0^{2}}}.
\end{align}
A direct check yields
\begin{align}
  F=H-TS-\Phi\,Q_{{e}}=G-\Phi\,Q_{{e}}.
\end{align}
This in turn justifies the explanation of $H$ as thermodynamic enthalpy.

Before proceeding, let us reveal a natural constraint of the black hole
solutions. From $f(r_0)=0$ we get $P=- \,{\frac {24\pi({r_{{0}}}^{2}c_{{1}}
+c_{{0}}r_{{0}}+d)}{{r_{{0}}}^{3}}}$. On the other hand, from eq.
(\ref{relation}) we get $d=-\,{\frac {{\epsilon}^{2}+{Q}^{2}-{3c_{{0}}}^{2}}
{c_{{1}}}}$. Inserting both into \eqref{temperature}, one can obtain
\begin{align*}
  T=\,{\frac {{\epsilon}^{2}+{Q}^{2}}{4c_{{1}}\pi \,{r_{{0}}}^{2}}}-\,{\frac
  { \left( c_{{0}}+r_{{0}}c_{{1}} \right) ^{2}}{4c_{{1}}\pi \,{r
_{{0}}}^{2}}},
\end{align*}
which leads to the constraint on $T$ and $r_0$
\begin{align}
  ({Q}^{2}+{\epsilon}^{2})-4\,c_{{1}}\pi \,r_0^{2}T\geq0.
  \label{tem}
\end{align}
This constraint implies an upper bound of $T$ at fixed horizon radius $r_0$,
or an upper bound of $r_0$ at fixed temperature $T$.

\section{$P-V$ criticality}

\subsection{Critical points}
To consider the $P-V$ criticality of the black hole, we should begin with the
equation of state (EOS) in $P-V$ plane at fixed conserved charges $Q$ and $c_1$.
The EOS arises from the expression \eqref{temperature} for the temperature $T$.
However, to use \eqref{temperature} as a reasonable EOS, we need to eliminate
the parameters $c_0$ and $d$. Assuming that both of these parameters are
nonzero. Then the condition $f(r_0)=0$ yields
\begin{align}
  c_0=-\frac{8}{3}\pi Pr_0^2-c_1r_0-\frac{d}{r_0}. \label{fr0}
\end{align}
Inserting \eqref{fr0} into eqs.(\ref{relation}) and (\ref{temperature})
respectively, one gets
\begin{align}
  &{\frac {64}{9}}\,{\pi }^{2}{P}^{2}r_0^{4}+ \frac{16}{3}\pi Pr_0
  \left(\,c_{{1}} \,r_0^{2}+\,d\right)+{c_{{1}}}^{2}r_0^{2}
  +{\frac {{d}^{2}}{r_0^{2}}} -{Q}^{2}-{\epsilon}^{2}-c_{{1}}d=0\label{eq1}\\
  &T=\frac{4}{3}Pr_0+\frac{c_1}{4\pi}-\frac{d}{4\pi r_0^2}\label{eq2}.
\end{align}
Then from eq.(\ref{eq2}), we find
\begin{align*}
  d=\frac{16}{3}\,\pi \,Pr_0^{3}-4\,T\pi \,r_0^{2}+c_{{1}}r_0^{2}.
\end{align*}
Inserting this into eq.(\ref{eq1}), we get the EOS
\begin{align*}
  64\,{\pi }^{2}r_0^{4}{P}^{2}- 16\pi\left( 4{\pi }T-\,c_{{1}}\right) r_0^{3}P
  +16\,{\pi }^{2}r_0^{2}{T}^{2}-4\pi \,c_{{1}}\,r_0^{2}T
  +c_1^{2}r_0^{2}-({Q}^{2}+{\epsilon}^{2})=0.
\end{align*}
Solving this equation for $P$ we get
\begin{align*}
  P=\,{\frac {T}{2r_{{0}}}}-\,{\frac {c_{{1}}r_{{0}} \pm \,\sqrt
{({Q}^{2}+{\epsilon}^{2})-4\,c_{{1}}\pi \,r_0^{2}T}}{8{\pi }
{r_{{0}}}^{2}}},
\end{align*}
and thanks to the condition (\ref{tem}), both branches of solutions should be
considered physical. Comparing the above expressions for the pressure with the
van der Waals equation
\begin{align*}
    P=\frac T{v-b}-\frac a{v^2}\simeq \frac T v+\frac{bT}{v^2}-\frac
    a{v^2}+O(v^{-3}),
\end{align*}
we are tempted to use the variable
\begin{align}
  v=2r_0
\end{align}
as an effective specific volume for the black hole system under consideration.
Thus we rewrite the full EOS as
\begin{align}
&  4\,{\pi }^{2}{v}^{4}{P}^{2}- 2\left( 4{\pi }T-\,c_{{1}}\right)\pi
  \,{v}^{3} P+4\,{\pi }^{2}{v}^{2}{T}^{2}-c_{{1}}\pi \,{v}^{2}T
  +\frac{\,c_1^{2}{v}^{2}}{4}-({Q}^{2}+{\epsilon}^{2})=0.
\label{EOS}
\end{align}

The critical points $(P_c,v_c,T_c)$ results from the conditions
\begin{align}
  &\frac{\partial P}{\partial v}|_{v=v_c,T=T_c}=0, \label{p1}\\
  &\frac{\partial^2 P}{\partial v^2}|_{v=v_c,T=T_c}=0, \label{p2}\\
  &P_c=P|_{v=v_c,T=T_c}. \label{p3}
\end{align}
The partial derivative in \eqref{p1} and \eqref{p2} can be evaluated directly
using \eqref{EOS}, which read
\begin{align}
  &\frac{\partial P}{\partial v}=-\,{\frac {2P}{v}}
  +\,{\frac {4\,T\pi-c_{{1}} }{4\pi \,{v}^{2}}}
  +{\frac {c_{{1}}T}{{v}^{2} \left(4\,T\pi -c_{{1}}- 4\,\pi \,Pv \right) }}
  \label{f1}\\
  &\frac{\partial^2 P}{\partial v^2}
  =\,\frac {6P}{{v}^{2}}-\frac {4\,T\pi-c_{{1}} }{{v}^{3}\pi }
  -\,\frac {3c_{{1}}T}{ \left(4\,T\pi -c_{{1}} -4\,\pi \,Pv \right) {v}^{3}}
  +\,\frac {4\pi \,{T}^{2}c_1^{2}}{ \left( 4\,T\pi -c_{{1}}
  -4\,\pi \,Pv\right) ^{3}{v}^{3}}.
\end{align}
Solving these two equations and substituting into \eqref{p3}, we get two sets
of critical point parameters as follows.
\begin{enumerate}
  \item When $c_1>0$,
\begin{align*}
  &v_c={\frac {\sqrt {X \left( {Q}^{2}+{\epsilon}^{2} \right) }}{c_{{1}}}},
  \quad T_c=\,{\frac { \left( 3\,X-8 \right) c_{{1}}}{4\pi \,
  \left( 27\,X-8 \right) }},
  \quad
  P_c=\,{\frac {-c_1^{2} \left( 3\,X-8 \right) }{16\pi \,
  \sqrt {X \left( {Q}^{2}+{\epsilon}^{2} \right) }X}};
\end{align*}

  \item When $c_1<0$,
\begin{align*}
  v_c={\frac {\sqrt {X \left( {Q}^{2}+{\epsilon}^{2} \right) }}{-c_{{1}}}},
  \quad
  T_c=\,{\frac { \left( 3\,X-8 \right) c_{{1}}}{4\pi \,
  \left( 27\,X-8 \right) }},
  \quad
  P_c=\,{\frac {c_1^{2} \left( 3\,X-8 \right) }
  {16\pi \,\sqrt {X\left( {Q}^{2}+{\epsilon}^{2} \right) }X}}.
\end{align*}
\end{enumerate}
In the above, $X$ can take two discrete values $X_1$ or $X_2$:
\begin{align}
  X_{1}={\frac {40}{3}}+\frac{16}{3}\,\sqrt {6}\simeq26.3973,\\
  X_{2}={\frac {40}{3}}-\frac{16}{3}\,\sqrt {6}\simeq0.2694.
\end{align}
However, the physical critical point must obey the constraints
$P_c>0$, $r_c>0$ and $T_c>0$, which lead to
\begin{align}
  X<\frac{8}{27}, \quad\text{$c_1>0$}.
\end{align}
This exclude the choice $c_1<0$ and $X=X_1$, and we are left with a single
critical point which is characterized by the parameters
\begin{align}
  v_c={\frac {\sqrt {X_2 \left( {Q}^{2}+{\epsilon}^{2} \right) }}{c_{{1}}}},
  \quad
  T_c=\,{\frac { \left( 8- 3\,X_2 \right) c_{{1}}}
  {4\pi \, \left( 8- 27\,X_2\right) }},
  \quad
  P_c=\,{\frac {c_1^{2} \left( 8- 3\,X_2 \right) }
  {16\pi \,\sqrt {X_2\left( {Q}^{2}+{\epsilon}^{2} \right) }X_2}},
  \label{cri}
\end{align}
in which $c_1> 0$. These parameters satisfy the relation
\begin{align}
  \frac{P_cv_c}{T_c}=\,{\frac {8-27\,X_{{2}}}{4X_{{2}}}}\simeq 0.6742,
\label{PvT}
\end{align}
which gives a pure number and is independent of all parameters. Actually, this relation is very similar to the one of the van der Waals system, which behaviors as $\frac{P_cv_c}{T_c}=\frac{3}{8}$ at its critical point. However, if one replaces $v$ in Eq.(\ref{PvT}) by the thermodynamic volume $V$, the result of this relation will be no longer independent of parameters. In this sense, it is more natural to consider $v$ as the specific volume instead of the thermodynamic volume $V$. Note that the
existence of critical point also exclude the possibility of $c_1=0$. Also,
the square sum of the charge $Q$ and the signature $\epsilon$ must also be
nonzero. Therefore, there is neither need to distinguish the case $Q=0$ from
the $Q\neq0$ cases (as long as $\epsilon \neq0$) nor need to distinguish the
$\epsilon =0$ case from the $\epsilon =\pm1$ cases (as long as $Q\neq 0$).

One may also be curious about the cases with $d=0$ (the BPS black hole) or
$c_0=0$, both of which will result in an EOS of ideal gas after considering eq.
(\ref{temperature}) directly. Thus they are all out of our discussion.
Another degenerated case $c_1=0,Q=0$ corresponds to the Schwarzschild-AdS
black hole. In this case one can never find physical critical points as is known
in \cite{Kubiznak:2012wp}.

\begin{figure}[h!]
\begin{center}
\includegraphics[width=0.45\textwidth]{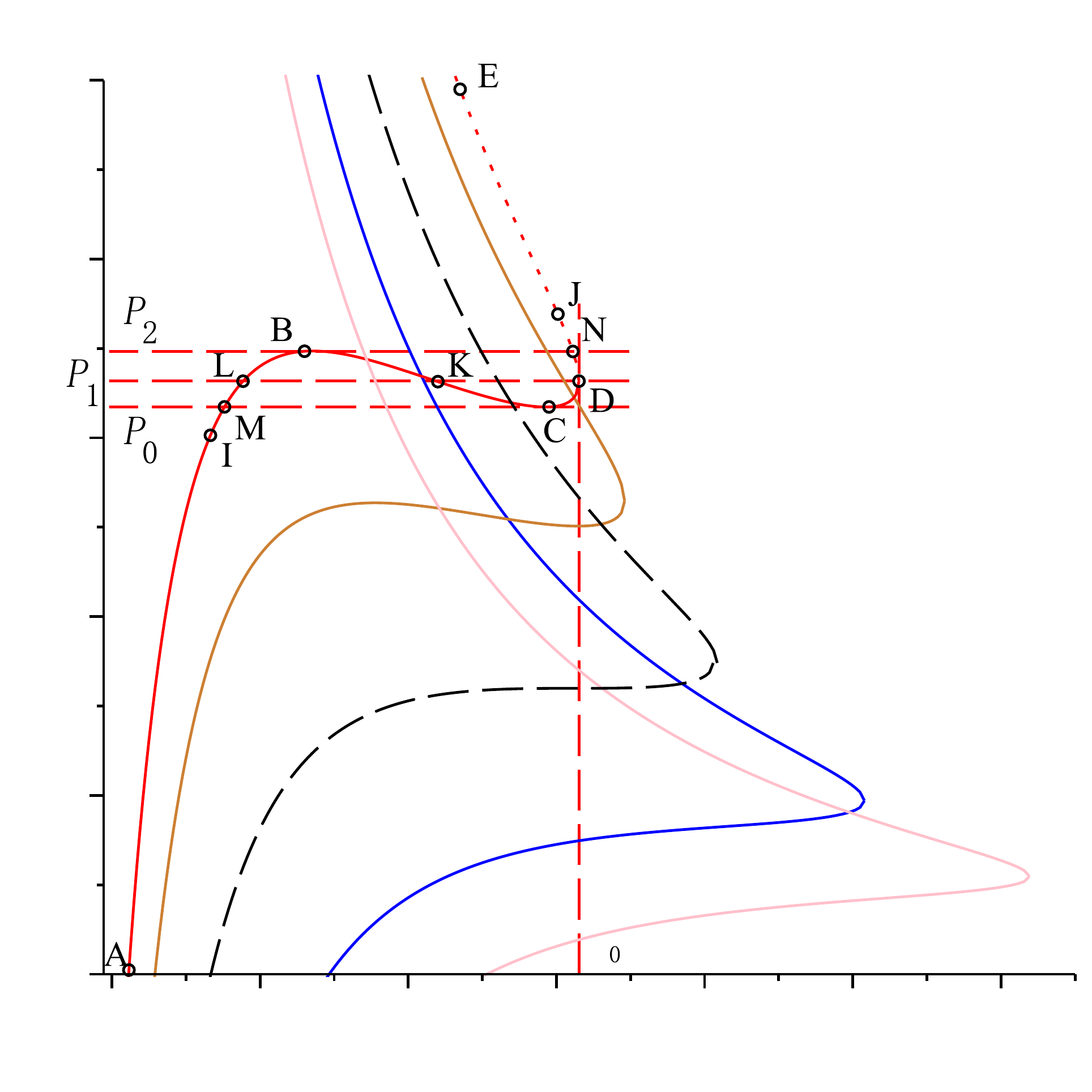}
\includegraphics[width=0.45\textwidth]{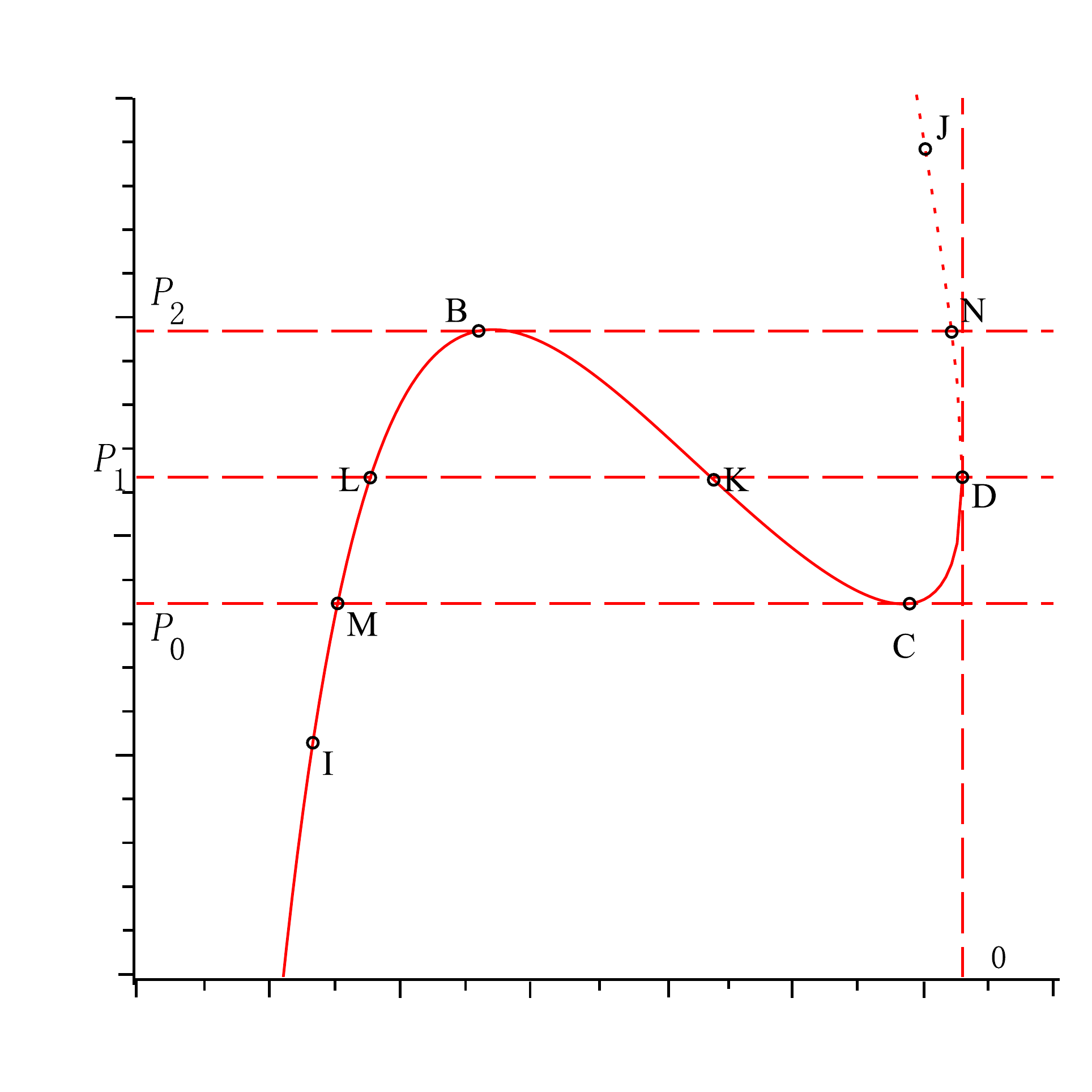}
\caption{The isotherms: on the left plot, at the intersection of each isotherms
with the horizontal axes, the temperature decreases from left to right. The
specific volume on each isotherm (horizon radius) has an upper
bound. The dashed curve corresponds to critical temperature. }
\label{Pv}
\end{center}
\end{figure}

The isothermal plots at generic parameters $Q, c_1$ are depicted in Fig.\ref{Pv}
(the right plot is a magnification of a single isotherm at the temperature
$T=1.5T_c$). While creating the plots, we take $\frac{\sqrt{Q^2+\epsilon^2}}
{c_1}, c_1$ and $\frac{c_1^2}{\sqrt{Q^2+\epsilon^2}}$ respectively as units for
$v_c, T_c$ and $P_c$. The pressure corresponding to the extremal specific
volume is denoted $P_1$.

It can be seen that on each isotherm there is an upper bound for the specific
volume (black hole radius) $v_{\mathrm{ex}}$. For $v<v_{\mathrm{ex}}$, the
isotherm can be subdivided into two segments, i.e. the lower branch and the
upper branch, which reflect the double-valuedness of the pressure. The
difference between the $T<T_c$ and $T>T_c$ curves lies in that, each branch
of the isotherm in the former case is monotonic with respect to the special volume $v$, while the lower branch
in the latter case is non-monotonic. Consequently phase transitions will occur
only in the $T>T_c$ regime.

Another way of subdividing the isotherms is according to the sign of
$\left(\frac{\partial P}{\partial v}\right)_T$, which is inversely
proportional to the isothermal compressibility
\[
\alpha \equiv \frac{1}{v}\left(\frac{\partial v}{\partial P}\right)_T.
\]
According to the sign of the
isothermal compressibility, each isotherm with $T>T_c$ can be subdivided
into 4 segments, two with positive isothermal compressibility and two with
negative isothermal compressibility.

Let us take a closer look at the
magnified plot given on the right diagram in Fig.\ref{Pv}. This is a curve
corresponding to isotherm with $T=1.5T_c$. The lower and upper branches of the
curve is joined together at the point D which corresponds to the extremal
specific volume $v_{\mathrm{ex}}$ and the pressure $P=P_1$.
On the lower branch (plotted in solid line) one can see that there is a local
maximum $P_2$ and local minimum $P_0$ for $P$, the corresponding points on the
isotherm are marked with B and C respectively. The 4 segments of the isotherm
are $\overarc{IB}, \overarc{BC}, \overarc{CD}$ and $\overarc{DJ}$
respectively. Among these, $\overarc{IB}$ and $\overarc{CD}$ have negative
isothermal compressibilities which imply that black hole states falling in
these segments may be unstable.

The right plots in Fig.\ref{Pv} is typical for all  $T>T_c$. At such
temperatures, we can subdivide the range of the pressure into 4 regimes:
$P < P_0$, $P_0 \leq P< P_1$, $P_1 \leq P < P_2$ and $P \geq P_2$. When
$P< P_0$, there is a single unstable black hole phase represented by the segment
$\overarc{IM}$. Due to its unstable nature, a perturbative increase in the
pressure would result in an increase of the black hole radius until it reaches
the state M, then through a phase transition it will enter the second regime for
the pressure, $P_0 \leq P< P_1$. In this second regime for the pressure, there are
3 different black hole states for a single pressure value, among these,
only the one with intermediate sized specific volume (i.e. the state lying on
the segment $\overarc{KC}$) is stable. Therefore, as the pressure increases,
the black hole state will evolve from the state C until the state K is reached
and then the pressure enters the next regime,  $P_1 \leq P < P_2$. There are
still 3 black hole states at each pressure values in this regime, however,
two of these have positive isothermal compressibility (i.e. the states lying on
the $\overarc{KB}$ and $\overarc{DN}$ segments), so it is hard to tell
which is more stable by looking at the EOS alone. If the pressure enters the
fourth regime, $P \geq P_2$, then there is only a single stable phase
which corresponds to states on the segment $\overarc{NJ}$.
From the above analysis, it is clear that there are two possible
phase transition points, the first one occurs at $P=P_0$, where the black hole
will most probably transit from the state $M$ to the state $C$ which is more
stable upon perturbation. The second phase transition point occurs
either at $P=P_1$ (if the Gibbs free energy on the segment $\overarc{KB}$
is higher than that on $\overarc{DN}$)
or at $P=P_2$ (if the Gibbs free energy on the segment $\overarc{KB}$ is lower
than that on $\overarc{DN}$). In the next subsection,
it will be clear that the Gibbs free energy on the segment $\overarc{KB}$ is
always lower than it is on $\overarc{DN}$, so the second phase transition point
occurs at $P=P_2$, where the black hole picks the state B instead of N,
because the state $B$  has lower Gibbs free energy than the state $N$.

\subsection{Gibbs free energy and the zeroth order phase transitions}

In order to have a further look at the critical points, we need to plot
the Gibbs free energy versus pressure at fixed temperature.
The Gibbs free energy $G$ as a function of $T$ and $P$ is quite complicated and
cannot be given explicitly with ease. So we will try to present the
$G-P$ relationship in terms of a pair of parametric equations.

\begin{figure}[h!]
\begin{center}
\includegraphics[width=0.45\textwidth]{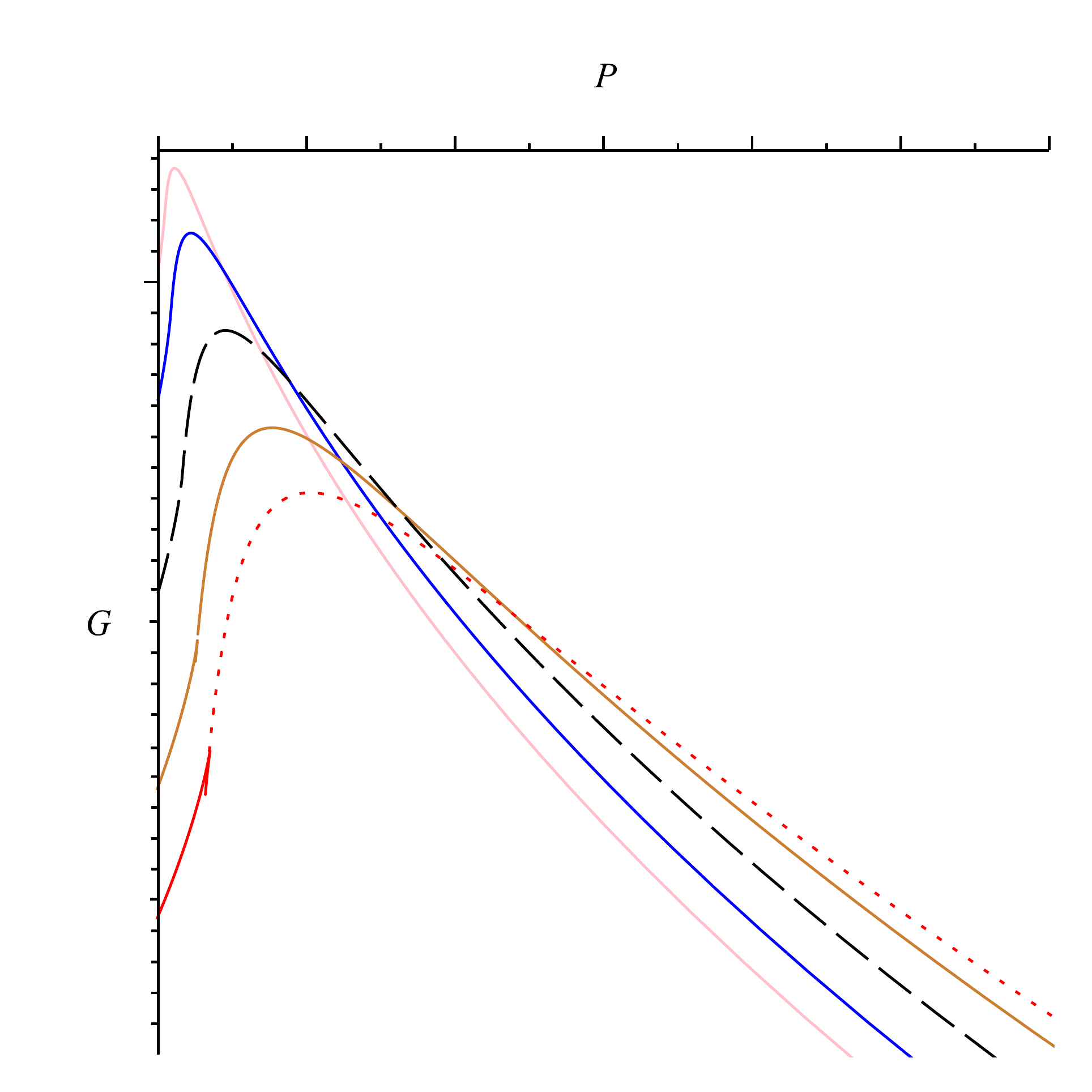}
\includegraphics[width=0.45\textwidth]{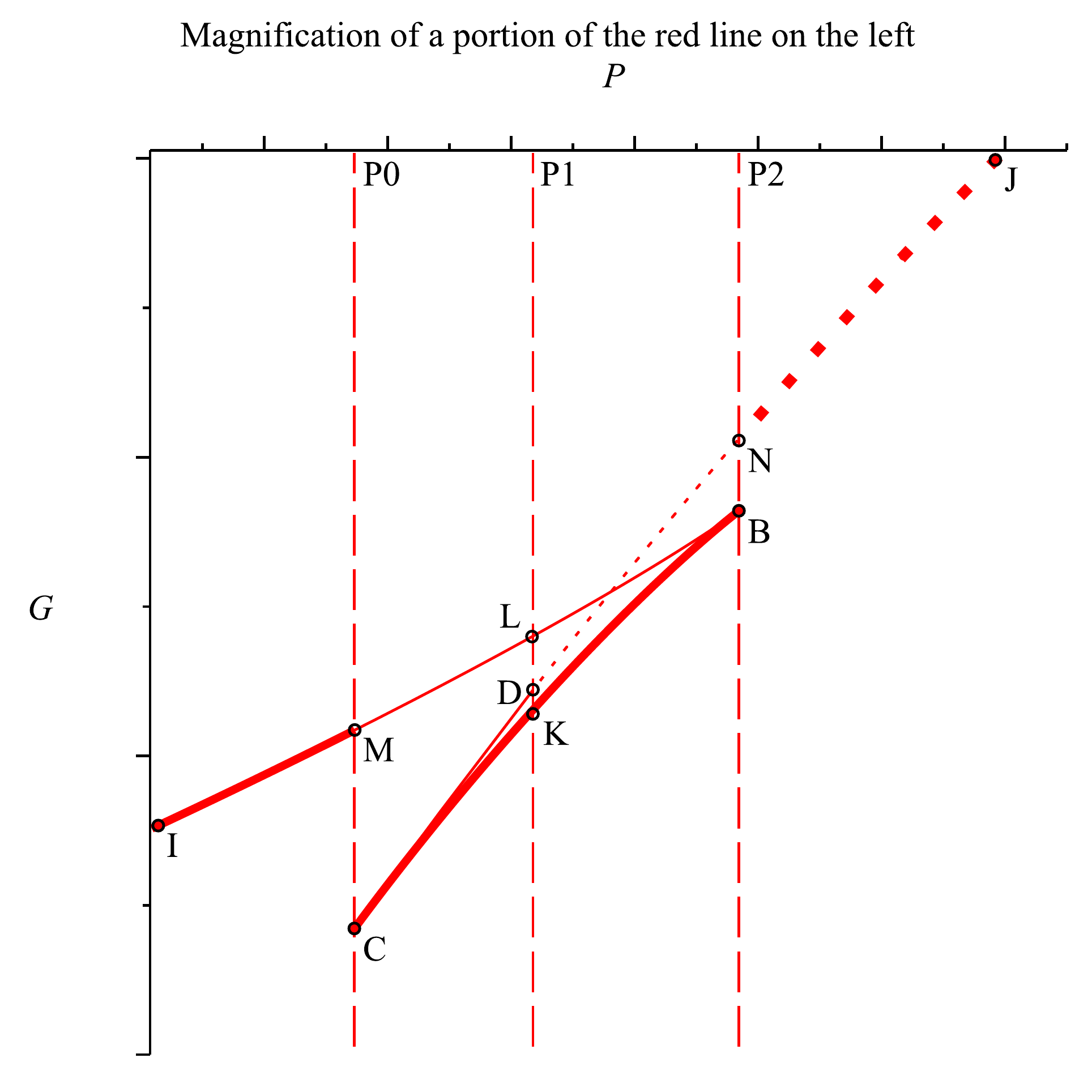}
\caption{The Gibbs free energy $G$ versus $P$ at constant temperatures.
On the intersections with vertical axes,
the temperature decreases from bottom to top, and the temperatures for each
curve are in one to one correspondence to the
isotherms given in Fig.\ref{Pv}.
The right plot is a magnification of the lower-left corner of the (red) line with the highest temperature in the left plot. The marked points are
the same as in the right plot of Fig.\ref{Pv}. The global minimum of the Gibbs free energy in the right plot is highlighted by the thick line.}
\label{GP}
\end{center}
\end{figure}

First we can  invert the relation \eqref{eq2} to get $P$ as a function of
$T$ and other parameters. Eliminating $c_0$ in the resulting expression by
use of \eqref{fr0}, we get
\begin{align}
&P(v,d)=\,{\frac {3d}{2\pi \,{v}^{3}}}
+\,{\frac {3(4\,\pi \,T-c_{{1}})}{8v \pi }}, \label{Pveq}
\end{align}
where we have also replaced $r_0$ by $v/2$. Inserting this equation together
with \eqref{fr0} into \eqref{Gibbs1}, we get
\begin{align}
&G(v,d)=\,{\frac { \left( 4\,\pi \,T-c_{{1}}\right) d}{12v\pi }}
  -\,{\frac { \left( 4\,\pi \,T+c_{{1}} \right) ^{2}v}{144\pi }}
  -\,{\frac {\epsilon\, \left( 4\,\pi \,T+c_{{1}} \right) }{12\pi }}
  -\,{\frac {2({Q}^{2}+{\epsilon}^{2})}{9v\pi }}. \label{Gv}
\end{align}
In \eqref{Pv} and \eqref{Gv}, $d$ cannot be taken as a free
parameter, because there is an extra constraint (\ref{eq1}), which can be
rewritten as
\begin{align}
\,{\frac {9{d}^{2}}{{v}^{2}}}
+ \frac{3d}{2}\left(4\,\pi \,T-\,c_{{1}} \right)
+\frac{\,{v}^{2}}{16} \left( 4\,\pi \,T+c_{{1}} \right)^{2}
-({Q}^{2}+{\epsilon}^{2})=0.
\end{align}
This quadratic algebraic equation gives two solutions for $d$ as a function of
$T,v$. Picking each solution and substituting into \eqref{Pveq} and \eqref{Gv},
the resulting pair of equations can be taken as parametric definition for the
function $G(P,T)$ for constant $T$. So, we can easily plot the $G-P$ diagram
at constant $T$, which is presented in Fig.\ref{GP}.

It can be seen in Fig.\ref{GP} that, for all $T>T_c$, a downcast
swallow tail appears on each $G-P$ curve.
From $T=T_c$ and downwards, the swallow tail disappears, with $T=T_c$
corresponding to the critical point. This is different from that of a Van der
Waals liquid-gas system, where $T<T_c$ is required for a phase equilibrium. The
same phenomenon has also been observed in the study of
criticality associated with the GB coupling constant
\cite{Xu:2013zea} in Gauss-Bonnet gravity.

The downcast swallow tail is different from the usual upcast one, which
corresponds to first order phase transitions \cite{Kubiznak:2012wp}.
On the magnified plot given on the right plot of Fig.\ref{GP}, it is clear that
the Gibbs free energy on the segment $\overarc{CKB}$ is lower than
that on $\overarc{CDN}$ and $\overarc{BLM}$. At $P=P_0$ and $P=P_2$, there
exist discontinuities for the Gibbs free energy, indicating that there are
zeroth order phase transitions at these two particular pressures.

\begin{figure}[h!]
\begin{center}
\includegraphics[width=0.45\textwidth]{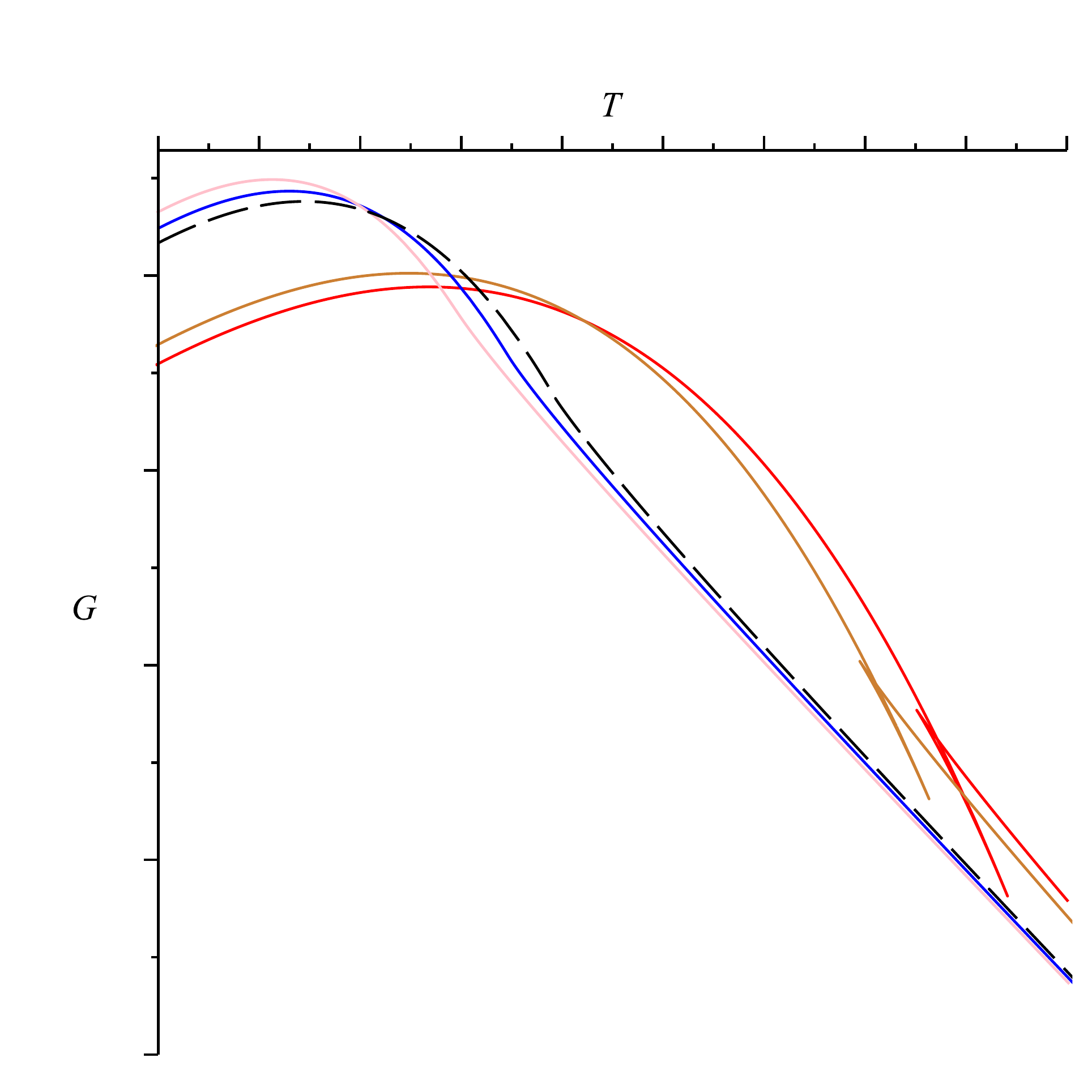}
\includegraphics[width=0.45\textwidth]{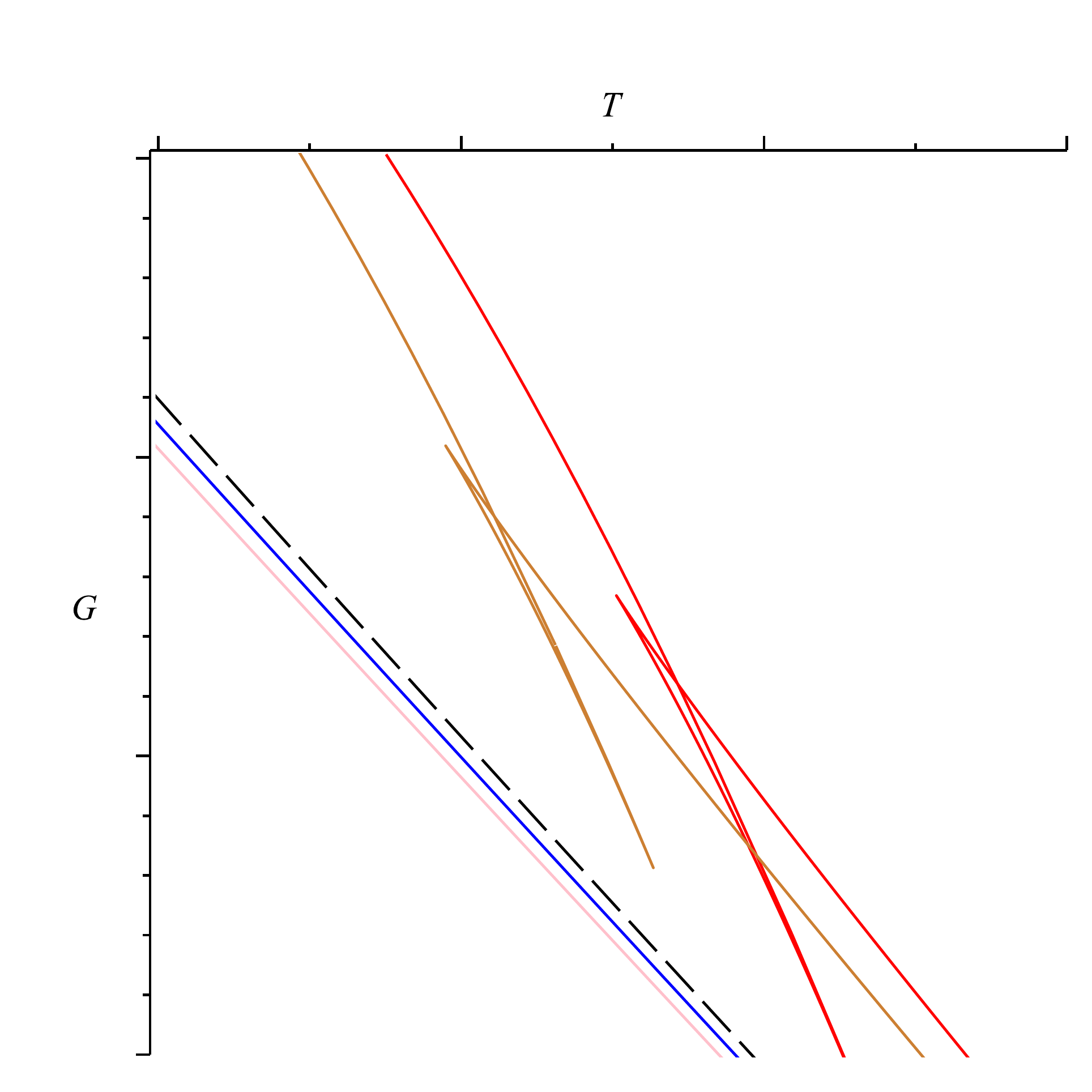}
\caption{The Gibbs free energy versus temperature at constant pressure: the
curves on the upper-right has higher pressure}
\label{GT}
\end{center}
\end{figure}

If one follows the red curve given in the left plot of Fig.\ref{GP},
it would be clear that at sufficiently high pressure, the Gibbs free energy
on the dotted segment would eventually become lower than it is on the solid
segment. This implies that the thermodynamics of the system favors large
pressure states. On the other hand, if one looks at the Gibbs free energy
versus temperature plots at fixed pressure (Fig.\ref{GT}), one would see
the similar downcast swallow tail at $P>P_c$. Moreover, following a single
$G-T$ curve reveals that the system favors high temperature states if the
pressure is kept fixed.

\section{Concluding remarks}

In this paper, we considered the $P-V$ criticality of static charged AdS
black holes in conformal gravity. Unlike the cases of Einstein gravity,
the cosmological constant arises as an integration constant in conformal
gravity, making the analysis for $P-V$ criticality more self-contained,
e.g. without need to consider systems with different actions.

The thermodynamics in the extended phase space for black hole in conformal
gravity possesses several unusual features:
\begin{itemize}
\item there exists only one critical point but there are two phase transition
points, both of which corresponds to zeroth order phase transitions;

\item there is no first order phase transition in the system;

\item the phase transition can occur only when $T>T_c$ (or $P>P_c$) but not the
other way round;

\item at fixed $T>T_c$, the system favors large pressure states, whilst at
fixed $P>P_c$, the system favors high temperature states.
\end{itemize}
We do not know of any other black hole or ordinary matter systems which exhibit
similar thermodynamic behaviors. It would be interesting to find other examples
which yield similar behaviors, because otherwise the system under study
would seem to be too bizarre to understand.

\section*{Acknowledgements}
We would like to thank Hong Lu for suggesting this research.

\providecommand{\href}[2]{#2}\begingroup
\footnotesize\itemsep=0pt
\providecommand{\eprint}[2][]{\href{http://arxiv.org/abs/#2}{arXiv:#2}}

\end{document}